\documentclass[letterpaper,12pt]{article}
\usepackage[hmargin=1in,vmargin=1in]{geometry}
\usepackage{natbib}
\usepackage{setspace,amssymb,amsmath,graphicx,wasysym,fancyhdr}
\usepackage{verbatim}
\usepackage[labelfont=bf]{caption}

\begin{document}

\title{Dimensional and statistical foundations for accumulated damage models}
\author{Samuel W K Wong$^1$ and James V Zidek$^2$\\
$^1$  Department of Statistics, University of Florida, Gainesville, FL\\
$^2$  Department of Statistics, University of British Columbia, Vancouver, BC\
}

\maketitle{}

\begin{abstract}
This paper develops a framework for creating damage accumulation models for engineered wood products by invoking the classical theory of non--dimensionalization. The result is a general class of such models. Both the US and Canadian damage accumulation models are revisited. It is shown how the former may be generalized within that framework while deficiencies are discovered in the latter and overcome.  Use of modern Bayesian statistical methods for estimating the parameters in these models is proposed along with an illustrative application of these methods to a ramp load dataset.
\end{abstract}

\newpage
\section{Introduction}\label{sec:introduction}
The reliability of manufactured lumber products used in structural engineering applications is assured by their design values. These values would be relatively easy to specify for short term loadings, for example in terms of estimated fifth percentiles of their breaking strengths under such loads.  But these design values must also account for the combination of short and long term future dead and live loads they must sustain. For this purpose the theory of accumulated damage models (ADMs) was developed \citep[for a review see e.g.,][]{hoffmeyer2007duration,Vincent2011}.

The seminal paper of Lyman Wood \citep{wood1951relation} played a key role in that development, by showing that the strength of lumber is a relative thing -- it depends on how loads are applied. His empirical assessments showed conclusively that the load at failure will be much higher when that load is applied at a rapidly increasing rate as compared with a slowly increasing rate.  This is called the duration--of--load effect and Wood quantified it in the so--called Madison Curve \citep{cai2016thesis, cai2016damage}. But these empirical assessments did not enable the strength of lumber under future loadings to be characterized. For that purpose models were needed.

ADMs were developed to meet that need. These models are parametric functions over time of the future stress loading profile $\tau(t),t \geq 0$; the damage accumulated over a long future could then be predicted using a $\tau(t)$ that reflected the types of loads, e.g.~snow, that might occur. Their ingenuity derived from the feasibility of estimating the model parameters from data obtained from  accelerated testing.  In other words, in laboratory experiments $\tau(t)$ could be chosen to ensure failure in a realistic time frame, e.g.~a ramp load test of duration about 1 minute to yield the short-term strength $\tau_s$ of a piece of lumber.  That short-term strength $\tau_s$ could be treated as a  property of a piece of lumber randomly selected from any given  in--grade population of interest. Or it could, depending on the context, be treated as a fixed parameter of that population. But in either case, its role could be expressed very simply in ADMs through the unitless stress ratio $\sigma(t) = \tau(t)/\tau_s$---the impact on a random piece of lumber of a general load profile at time $t$ would be calibrated by the multiplicative rescaling factor $\tau_s^{-1}$. That stress ratio became a fundamental determinant in models of the rate at which damage to a piece of lumber accumulated over time as the future load is applied.

A general form for an ADM is given in \citet{rosowsky2002another}, which states that damage accumulates at a rate determined by
\begin{equation}\label{eq:geneqn}
\frac{d\alpha(t)}{d t} = g(\alpha(t), \sigma(t), \boldsymbol{\nu}),	\end{equation}
where $\alpha(t)$ is the damage accumulated up to time $t$,  $g$ is a function to be specified,  $\boldsymbol{\nu}$ is a vector of parameters, and $\sigma(t)$ is the applied stress ratio at time $t$ as defined above. The general model in Equation (\ref{eq:geneqn}) has been extended by \citet{kohler2002probabilistic} to include additional model parameters $\boldsymbol{\xi}$ to be fitted using experimental data, where the form of the model is derived from engineering theory, e.g.~crack formation theory.  While the focus here is models based on Equation (\ref{eq:geneqn}), the approaches discussed are also applicable to the more general models.

The accumulated damage $0\leq \alpha(t)\leq 1$ is a non-decreasing function of $t$. At time $0$, $\alpha(0) = 0$, as no damage will yet have occurred. It is assumed that the loading is sufficiently large as to cause failure at a finite random time $t = T_l$, at which time accumulation of damage is complete and that $\alpha$ is scaled so that $\alpha(T_l) = 1$. Note that the ``accumulated damage'' $\alpha(t)$ is  latent---it is not an observable characteristic of the piece of lumber. Instead it provides a framework on which to hang the various elements of the model.

An important feature of the models is their temporal scales.  The values of $\tau_s$, $\sigma(t)$, and $\tau(t)$ at any fixed point in time do not depend on the units of measurement chosen for $t$ (e.g.,~whether time is measured in minutes or seconds).  But the same cannot be said of the rate at which damage is accumulated as specified by Equation (\ref{eq:geneqn}). This rate must by definition depend on the unit scale adopted for the dimension of time. In fact the dimensions and the scales on which they are measured are a fundamental aspect of any general theory for a natural phenomenon.  However the ADMs have been inconsistent in the way time and other parameters have been incorporated in the model, leading to unrecognized technical anomalies \citep{cai2016thesis, cai2016damage}.  Thus the first major result of this paper is to develop a new approach to constructing ADMs by non--dimensionalizing the problem.

Fitting ADMs has proven challenging, because Equation \ref{eq:geneqn}  does not readily yield a likelihood function, which is the cornerstone of conventional statistical approaches for estimating model parameters. Instead various complicated methods for estimating those parameters were developed \citep{foschiyao1986dol,gerhards1987cumulative}, although statistical properties such as their standard errors are difficult to assess.  Thus a second major contribution of this paper is a new and principled statistical foundation based on the use of Bayesian methods to incorporate both randomness between specimens as well as model error; while much more computationally intensive, the abundance of modern computing power makes their applicability to these models now feasible. This new approach necessitated the development of code for implementation over a large cluster of CPU cores, used in this paper to compare two well--known ADMs after appropriate non-dimensionalization.
	
To summarize, Section \ref{sec:model} shows in detail how one may develop a model for a natural process, in this case damage accumulation, by first non--dimensionalizing to canonical form, thus bypassing the need for scales of measurement. Then variations of models are developed that have been of fundamental importance in the development of design values that account for uncertainties in future loading profiles.  Section \ref{sec:experiment} describes the data obtained from a ramp-load experiment in the FPInnovations testing laboratory.  Section \ref{sec:results} provides a novel illustrative application where models are fitted and compared using the Bayesian statistical methods described in this paper.  The paper concludes with a brief discussion in Section \ref{sec:conclusions}.

\section{New models through dimensional analysis}\label{sec:model}
This section focuses on the concept of dimension as it relates to ADMs. The following is an illustrative example.
\vskip .1in
{\bf Example 2.1} Suppose two scientists $S_{min}$ and $S_{hr}$ working respectively on time scales of minutes $m$ and hours $h$, are engaged on a damage modeling project. Their respective objectives are accumulated damage models, $\alpha_{min}$ and $\alpha_{hr}$.  They know the stress ratios, $\sigma_{min} (m) $ and $\sigma_{hr} (h) $, respectively and also that $\sigma_{hr} (h) =\sigma_{min} (60h) $ since they are working on the same project. They also know in the end that
their times to failure must be the same $M = 60H$, that $\alpha_{hr} (h) = \alpha_{min} (60h),~h\geq 0$ and that $1=\alpha_{hr} (H) = \alpha_{min} (60H),~h\geq 0$.

  They now proceed to solve Equation (\ref{eq:geneqn}) to get
\begin{eqnarray}\label{eq:hrmineqn}
  \alpha_{min}(m) &=& \int_0^m g(\alpha(m^\prime), \sigma(m^\prime), \boldsymbol{\nu})dm^\prime \nonumber \\
  \alpha_{hr}(h) &=& \int_0^h g(\alpha(h^\prime), \sigma(h^\prime), \boldsymbol{\nu}) \, dh^\prime.
\end{eqnarray}

To check her results, $S_{hr}$ does further analysis and finds
\begin{eqnarray*}
	\alpha_{hr}(h) &=& \int_0^h g(\alpha_{hr}(h^\prime), \sigma_{hr}(h^\prime), \boldsymbol{\nu})dh^\prime \\
	&=& \int_0^h g(\alpha_{min}(60 h^\prime), \sigma_{min}(60h^\prime),\boldsymbol{\nu})dh^\prime.	
\end{eqnarray*}
Changing the variable of integration,
$m^\prime = 60 h^\prime$, yields
\begin{eqnarray}\nonumber
	\alpha_{hr}(h) &=&  \frac{
	\int_0^{60h} g(\alpha_{hr}(m^\prime), \sigma_{hr}(m^\prime), \boldsymbol{\nu})dm^\prime}{60} = \frac{\alpha_{min}(60h)}{60}, 
\end{eqnarray}
leading to a contradiction since $1 =\alpha_{hr} (H) = \alpha_{min} (60H)/60 = \alpha_{min} (M)/60 = 1/60$.
\vskip .1in

This contradiction in Example 2.1 could be resolved were there an international standard unit for time denoted by $[t]$ and by expressing $t$ as a number, $\{t\}$ of standard units. That approach has been used to define the index $pH$ of acidity of an aqueous solution: it is defined by $pH= -\log_{10}\{a_H^+\}$ where $\{a_H^+\}$ is the number of internationally agreed on units of its hydrogen ion concentration.  Thus $pH$ becomes unitless and $pH = 7 $ for example, always represents the acidity of distilled water. This approach also bypasses another problem, that as a transcendental function, the logarithm $x\rightarrow\log (x)$ cannot be applied to $ x $ unless it is unitless \citep{matta2010can}.

However this technical ``fix'' does not seem satisfactory for modelling the strength properties of lumber. Different time units may be preferable to others in certain contexts, in particular since ADMs are used in both short-term testing and long-term reliability.  More importantly, if $\{T_l\}$ were the random time to failure on the standardized scale, results could not be interpreted on  another time scale, e.g.~`hours'.  As the above analysis shows, different results would be obtained if the model had been built say on the hourly time scale, and now everything including the fitted model parameters were to be synchronized to that standard unit of time.

An alternative solution would include a time rescaling factor.  The following example illustrates that approach using a special case of model (\ref{eq:geneqn}).
\vskip .1in
{\bf Example 2.2:} As originally formulated the US accumulated damage model \citep{gerhards1987cumulative} is given by
\begin{equation}\label{eq:usmod}
\dot{\alpha}(t) = \exp{\{-A+ B\sigma(t)\}}
\end{equation}
where the `dot' means derivative with respect to time, and the parameter vector here is $\nu = (A,B)$.  This model cannot be correct as formulated, since the exponential function
$x\rightarrow\exp{(x)}$ cannot be applied to $ x $ unless the latter is unitless.   This can be corrected by defining
\begin{equation*}\label{eq:usmod1}
\dot{\alpha}(t) = C([t])~\exp{\{-A+B\sigma(t)\}}
\end{equation*}
where $C([t])$ depends on $t$ only through the units
$[t]^{-1}$ on the time scale on which it is measured, i.e.~the units of $C([t])$ are the inverse of the units of $t$.  Then Model (\ref{eq:usmod}) yields in this special case
\begin{eqnarray*}\nonumber
	\alpha_{hr}(h) &=& \int_0^h C([h^\prime]) \exp{\{-A+B\sigma_{hr}(h^\prime)\}}dh^\prime \\\label{eq:usmod2}
	&=& \int_0^h C([h^\prime]) \exp{\{-A+B\sigma_{min}(60h^\prime)\}}dh^\prime.	
\end{eqnarray*}
If the variables are transformed as above $m^\prime = 60h^\prime $, $C([h^\prime])
 \rightarrow C([m^\prime])60$ so that the $60$ cancels out in Equation (\ref{eq:hrmineqn}), thereby eliminating the inconsistency seen above.
\vskip .1in

However the approach illustrated in Example 2.2, proves impractical in cases that additionally have many parameters with associated units of measurement. Of greater concern then is the possibility that the model itself is dimensionally inconsistent, in which case it could not be said to represent a natural phenomenon \citep{shen2015nondimensionalization}. These considerations lead to the approach taken in this paper, of reducing the model to its canonical form by non--dimensionalizing it and hence eliminating the scale altogether and thus concerns about the units in which things are measured.  The approach, developed in the next subsection, shows one well--known ADM to have inconsistencies, that cannot be simply resolved as suggested above by including a scale parameter.

\subsection{Non--dimensionalizing models}\label{subsec:nondimen}
Although dimensional analysis has a long history \citep{bluman2013symmetries}, this paper focuses on the celebrated Buckingham
$\pi$ theorem \citep{buckingham1914physically}, which resolves the inconsistencies noted above.   That theorem assumes that the scientist has specified a meaningful and complete set of quantities (or variables) $Q_1,\dots, Q_n$ for the phenomenon under investigation.  The goal is a model that specifies their relationship:
\begin{equation}\label{eq:frelation}
f(Q_1,\dots, Q_n)=0.	
\end{equation}
The remarkable $\pi$ theorem shows that under mild conditions, this characterizing relationship can always be re--expressed in a simpler, dimensionless form through what Buckingham calls $\pi$ functions, which satisfy $F(\pi_1,\dots,\pi_p)=0$, where the $\pi_i = \Pi_{j=1}^n Q_j^{a_{ji}}$
are dimensionless and hence unitless. The $\{\pi\}$'s can thus be
considered to be the fundamental building blocks of the relationship expressed in
Equation (\ref{eq:frelation}).  Moreover estimating $F$ may be much simpler than estimating $f$ based on experimental data, since $p$ can sometimes be much smaller than $n$.

This section shows how the theorem can be applied, specifically to develop alternatives to well--known ADMs that ensure dimensional consistency. But the method can be applied more generally in developing engineering models with the potential benefit of simplifying the experiments needed to fit the core relationship amongst the quantities related by the models. One famous example, from fluid dynamics concerns the force $F$ on a body immersed in a fluid stream, which depends on the body's length $L$, fluid velocity $V$, fluid density $ \rho$, and fluid viscosity $ \mu$. Of interest is the relationship $g$
\[
F = g(L,V,\rho, \mu ).
\]
An experiment designed to estimate $g$ would be complex since all five of these quantities would seemingly need to vary. However the approach to be described below, when applied to this case shows that the fundamental relationship amongst these quantities actually involves just two quantities,  one being the dimensionless Reynolds number for the fluid, $Re = \rho VL/\mu$. More precisely
\[
F^* = g^*(Re),
\]
where $F^* = F/(\rho L^2V^2)$. A much simpler experiment yields an estimate $g^*$ from which the desired estimate of $g$ can be found.

The application in this paper has as a  primary goal, to relate the
rate at which the damage accumulation model
$\alpha(t)$ changes at
a time $t$, to other features of a randomly  chosen specimen
of an engineered wood product. Denote that change by $\dot{\alpha}(t) = \frac{\partial\alpha(t)}{\partial t}$.  Through its random quantities, that model represents the population from which that specimen is drawn. For those quantities, the models
that have already been proposed, such as those seen in the sequel, guided the selection of specific versions of
Equation (\ref{eq:frelation}).

In general the
dimensions of quantities are represented
by using square bracket notation. Thus
a quantity $Q$ would be written as
$Q=\{Q\}[Q]$ where
$[Q]$ is the dimension of $Q$ while $\{Q\}$ is
the number of units the quantity has in that
dimension.  In the physical sciences the
primary dimensions are time denoted by
$T$, mass $M$ and length $L$ (used for
any dimension of size including width, height,
thickness, etc.). Thus in practice
$t=\{t\}[t]$ where
$[t]=T$, the dimension being time. Once a dimension like time has been identified,
a scale has to be assigned according to
how that dimension is to be quantified or
measured.

A key element of an ADM is its rate
of change,
$Q_1 = \dot{\alpha}(t)$ with $[Q_1] = T^{-1}$.
Modellers (see for example \citet{foschiyao1986dol})
have assumed that it depends in a Markovian
way on the accumulated damage, i.e.~on
 $Q_2= \alpha(t)\in [0,1], [Q_2]=1^0$, a dimensionless quantity. The value $\alpha = 1$ is reached when the random specimen fails.  Note that
 $\dot{\alpha}(t)$ may depend on $t$ only indirectly, through some other quantity.  The rate of change at a specific time also depends on the stress $Q_3=\tau(t),~[Q_3] = FA^{-1}$
 where $A = L^2$ denotes the dimension of area and $F$ denotes
the dimension of force. As noted above, the
short term breaking strength
$Q_4 = \tau_s,~[Q_4] = FA^{-1}$, plays a key role.
It has generally been represented by the breaking strength under a ramp load test of short duration with a loading profile $\tau(t) = kt$
for a constant load rate $k,~[k]= FT^{-1}$, that is
\begin{equation}\label{eq:ramptest}
\tau_s = k T_s,	
\end{equation}
where $T_s,~[T_s]=T$ is the short term breaking time.
As applied to modelling the accumulated
damage, $T_s$ is a latent characteristic
of a piece of lumber or its corresponding population
parameter, whereas $k$ would be known.

Note that
Equation (\ref{eq:ramptest}) also holds approximately
under another
type of short-term test where it is the
deflection rate, not the load rate, which
is held constant. For completeness,
that type of ramp test will now be
described along with the $k$ involved, which
must now depend on the piece of lumber
(Conroy Lum, personal communication).
To illustrate the calculation of $k$ in a simple
case, suppose the piece is
anchored at its ends
in a bending machine. The span
or distance between the supports
is $L^*$. Two
downward acting loads $F^*/2$ are each applied
at equi--spaced points along the span.
In reaction, this induces upward acting loads $F^*/2$
at the ends of the span for a total of four
loads acting on the member.
Standard beam theory implies that at time
$t$ the maximum
deflection at mid span is
\begin{eqnarray}\label{eq:deflection}
	D(inches) & = & \frac{F^*L^*/3}{48EI}
	\left[3L^{*2} - 4\left({L^*\over3}\right)^2 \right] = \frac{23F^*L^{*3}}{1296EI}
\end{eqnarray}
where $E (psi)$ is the specimen-specific measure
of elasticity, and $I = bd^3/12$, the areal moment
of inertia, $b$ being the breadth of the
member and $d$ being its depth.
Equation (\ref{eq:deflection}) may thus
be simplified as
\begin{eqnarray*}\nonumber
	D(inches)
	& = & \frac{276F^*L^{*3}}{1296Ebd^3}.
\end{eqnarray*}

During the test  the force
 will dynamically increase
over time so that at time $t$
\begin{eqnarray*}\nonumber
	D(t)(inches)
	& = & \frac{276F^*(t)L^{*3}}{1296Ebd^3}  =  \frac{CF^*(t)}{E},
\end{eqnarray*}
for a constant $C>0$.  Thus requiring a constant deflection rate
$\dot{D}(t) = d$ implies
\begin{eqnarray}\label{eq:deflectionconstant}
	d & = & C \dot{F}^*(t)/E,
\end{eqnarray}
which means $F^*(t) = kt$
where
\begin{eqnarray}\label{eq:deflectionrateload}
	k & = & Ed/C.
\end{eqnarray}
Equation (\ref{eq:deflectionconstant}) shows that
if a constant deflection rate $d$ is to be maintained over time, the force $F^*(t)$ must be
adjusted to a higher value when $E$ is large than
when it is small.
In general the calculation above
would need to be adapted to the particular
test being used. But it does show that
$k$ can be calculated explicitly, knowing
$E$, so is not a random effect. It
also shows that the effect of $E$
is absorbed in $k$ so it need not
be included as a quantity in the model.
Thus in $\tau_s$ only the
time to failure $T_s$ (with $[T_s]=T$)
is random.

For more general testing scenarios that differ from the standard ramp load, it is not $T_s$ (with $[T_s]=T$) that is observed, but rather the time to failure under a given load profile $\tau(t)$ which shall be denoted by $T_l$.
When the accumulation of damage is complete,
the specimen fails and
$\alpha(T_l) = 1$. While failure time $T_l$ is clearly an important quantity, it is specimen-specific and derived from
$\alpha$ and hence need not explicitly be included in the model.  Instead, a reference level for time that is estimable from the experimental data might be used, for example the population average time of $T_l$ denoted by $\mu_l$. That feature is therefore included in the model as $Q_5 = \mu_l,~[Q_5] = T$.

The rate of change in $\tau(t)$,
 $Q_6 = \dot{\tau}(t),~[Q_6] = F(AT)^{-1}$,
has not been considered in previous models.
But for completeness, it is now shown
how it could be made part of the general framework.

 Finally there is the size of a specimen
 as a determinant of the rate at which
 damage is accumulated. Size would be
 characterized by a number of features,
depending on the nature of the
product.
For definiteness, assume just
three, $Q_7 = Width$, $Q_8=Thickness$ and
$Q_9 = Length$. Formally they all have
the dimension of length $L$. Thus for example, $[Thickness] = L$.  These features
will all be constants when interest focuses on a
specific size class.  But in the context of modelling
the full in--grade population based on
a random sample, these quantities
will vary and thus are included in the general model as well.

The above list of quantities with
 their units is summarized in Table \ref{table:quantities}.

 \begin{table}
\begin{center}
\begin{tabular}{l|cccccccccc}
\hline
  & $Q_1$ &$Q_2$ &$Q_3$& $Q_4$&$Q_5$ &$Q_6$ &$Q_7$&$Q_8$ &$Q_9$ \\
  \hline
Quantity    & $\dot{\alpha}(t)$
 & $\alpha(t)$
 & $\tau(t)$
 & $\tau_s$
 &  $\mu_l$
 &  $\dot{\tau}(t)$
 &  $ W $
 &  $ T$
 &  $ L$\\
Units
& $T^{-1}$
& $1^0$
& $FA^{-1}$
& $FA^{-1}$
& $T$
& $FA^{-1}$
& $L$
& $L$
& $L$ \\
\end{tabular}
\caption{The Q functions for accumulated damage models. Here $Width$, $Thickness$ and $Length$ are abbreviated by $W$, $T$ and $L$
for brevity.}\label{table:quantities}
 \end{center}
\end{table}

The $\pi$ theorem can be applied in various ways, depending on which
dimensions are chosen as the primary ones, and which the secondary. Note that in the summary above
only three primary or reference dimensions, $L$, $F$ and $T$ are manifest.
This implies there are just $3$ so--called ``repeating  quantities'' and $9-3=6$ $\pi$ functions. The repeating quantities cannot include the model's predictand $Q_1$. Previous work has shown $Q_4$ to be important as a baseline
measure of strength.  The average failure time of the population seems a good choice given its importance as a parameter.  $Q_9$ could well be chosen to represent the length $L$ group of quantities, especially if the population
specimens were of fixed length, but of varying
width and thickness.

These considerations suggest forming
the $\{\pi\}$ functions by first eliminating $Q_4$ as well as
$Q_9$ and then successively modifying $Q_1$,$Q_2$,$Q_3$,$Q_6$, $Q_7$ and $Q_8$. To illustrate the process, $Q_1$ is added to the repeating variables
$Q_4^b$ and $Q_8$ to form the first $\pi$ function as
\begin{eqnarray*}
	\pi_1 & \doteq & Q_1 Q_4^a Q_5^b Q_9^c
\end{eqnarray*}
with $a$,$b$, and $c$ chosen to make $\pi_1$ dimensionless.
This is interpreted in dimensional terms as
\begin{eqnarray*}
	 (T^{-1}) (FA^{-1})^a (T)^b (L)^c  &=& F^0 T^0 L^0
\end{eqnarray*}
giving $a=0$, $b = 1$ and $c=0$. Thus
\[
\pi_1 = Q_1 = \dot{\alpha(t)}\mu_l.
\]
Similarly,
\begin{eqnarray*}
	\pi_2 & \doteq & Q_2 Q_4^a Q_5^b Q_9^c~\textrm{,~and~hence,~the~restriction} \\
	(1^0)(FL^{-2})^a (T)^b (L)^c  &=& F^0T^0 L^0
\end{eqnarray*}
meaning that
\[
\pi_2 = Q_2 = \alpha(t);
\]
Continuing,
 \begin{eqnarray*}
	\pi_3 & \doteq & Q_3 Q_4^a Q_5^b Q_9^c \\
	(FL^{-2}) (FL^{-2})^a (T)^b (L)^c &=& F^0 T^0 L^0,
\end{eqnarray*}
 which yields
 \[
\pi_3 = Q_3Q_4^{-1} = \frac{\tau(t)}{\tau_s}.
\]
The remaining $\pi$ functions can be obtained in a similar fashion:
 \begin{eqnarray*}
 \pi_6 & \doteq & Q_6 Q_4^{-1} Q_5 = \frac{\dot{\tau}(t)\mu_l}{\tau_s} \\
 \pi_7 & \doteq & Q_6 Q_9^{-1} = \frac{Width}{Length}\\
\pi_8 & \doteq & Q_7 Q_9^{-1} = \frac{Thickness}{Length}.
\end{eqnarray*}

  Buckingham's theorem implies  $F(\pi_1,\dots,\pi_p)=0$, or
 \begin{equation}\label{eq:piversion}
 	\pi_1 = F^*(\pi_2,\pi_3, \pi_6,\pi_7,\pi_8),
 \end{equation}
 that is
  \begin{equation}\label{eq:piversion2}
 	\dot{\alpha}(t)\mu_l  = F^*\left(\alpha(t),
 	\frac{\tau(t)}{\tau_s}, \frac{\dot{\tau}(t)\mu_l}{\tau_s},\frac{Width}{Length},\frac{Thickness}{Length}\right).
 \end{equation}

 \textbf{Remarks:}
 \begin{enumerate}
 	\item This application of Buckingham's theory eliminates length as predictive of the rate of accumulative damage in agreement with the standard models like those in Sections \ref{US:Model} and \ref{sec:canadianmodel}. But those models unlike the ones proposed in this paper also exclude width and length. This may be reasonable in the case of short term (ramp) tests since the cross sectional area is already
		represented in the moment of areal inertia, that in turn, like the modulus of rupture, is absorbed in coefficient $k$ in Equation
		(\ref{eq:deflectionconstant}). But the rationale for this exclusion for
		an arbitrary loading curve $\tau(t)$ is unclear to these authors.
	\item Although Equation (\ref{eq:piversion2}) was developed with reference to a specific time point $t$, the same relationship holds for all $t\in [0,T_l]$ where $T_l$ denotes the time at which the specimen fails.  Hence the $\pi$ functions that are expressed as functions of $t$ are genuinely time--dependent.
	\item Equation (\ref{eq:piversion2}) provides a fundamental relationship amongst all the quantities in the characterizing relationship given in  Equation (\ref{eq:frelation}).  The functions $f$ and $F$ remain to be specified by some combination of scientific methods and experimental work. As they are models for a randomly selected specimen, they will be random.  Moreover they, like all models, will be inexact and hence require the inclusion of an uncertain model error; the Bayesian context of the paper requires they must be treated as random. Examples of ways of incorporating that uncertainty follow in the sequel.
 \end{enumerate}

 \subsection{The US Model}\label{US:Model}

 This section introduces a special case of the model in Equation (\ref{eq:piversion}), namely the one in Equation (\ref{eq:usmod}) for which  \begin{eqnarray}\nonumber
 	\pi_1(t) &=& F^*(\pi_3(t))\\
 	&=& \exp{\{-A + B \frac{\tau(t)}{\tau_s}\} }\label{eq:piversion3}.
 \end{eqnarray}
Here, $A$ and $B$ are random effects that reflect
residual model uncertainty since $\tau_s$ does not capture
all the variation from specimen--to--specimen.  This model is an amended version of the so--called US Model in that
now unlike before, the left hand side is dimensionless in agreement with the
right.

Integration yields
 \begin{eqnarray}\label{eq:usalpha}
 	\alpha(t) \mu_l
 	&=& \exp\{-A\}  \int_0^t \exp{\{B \frac{\tau(t^\prime)}{\tau_s}\} }dt^\prime.
 \end{eqnarray}

Specifically at the failure time $t = T_l$,
\begin{eqnarray}\label{eq:usalpha0}
 	\mu_l &=& \exp\{-A\} \int_0^{T_l} \exp{\{B \frac{\tau(t^\prime)}{\tau_s}\} }dt^\prime.
\end{eqnarray}
Observe that Equation (\ref{eq:piversion3}) implies $\pi_1(0) = \dot{\alpha}(0) \mu_l = \exp{\{-A\}}$, so using this in Equation (\ref{eq:usalpha0}) gives
$$
 	\dot{\alpha}(0) = \left[ \int_0^{T_l} \exp{\{B \frac{\tau(t^\prime)}{\tau_s}\} }dt^\prime \right]^{-1}.
$$
Then dividing by $\mu_l$, (\ref{eq:usalpha}) can be re-expressed as
 \begin{eqnarray}\label{eq:usalpha2}
 \alpha(t) &=& \dot{\alpha}(0) \int_0^t  \exp{\{B \frac{\tau(t^\prime)}{\tau_s}\} }dt^\prime \nonumber \\
 	&=&\frac{\int_0^t \exp{\{ B \frac{\tau(t^\prime)}{\tau_s}\} }dt^\prime}{ \int_0^{T_l} \exp{\{B \frac{\tau(t^\prime)}{\tau_s}\} }dt^\prime }\nonumber \\
 	&=&\frac{\int_0^{(t/\mu_l)} \exp{\{ B \frac{\tau(u\mu_l)}{\tau_s}\} }du}{\int_0^{(T_l/\mu_l)} \exp{\{ B \frac{\tau(u\mu_l)}{\tau_s}\} }du}
 \end{eqnarray}
by the change of variables $u = t^\prime/\mu_l$, which is exactly what was obtained in Section \ref{sec:model} by the {\it ad hoc} approach taken there.

 For the special case of a ramp load test the substitutions $T_l = T_s$, $\mu_l = \mu_s$, and $\tau(t) = kt$ can be made, where $k$ is the known loading rate (which
may vary between specimens if a constant deflection rate is maintained), $[k] = FL^{-2}T^{-1}$, $\tau_s = k T_s$, and $\mu_s$ is the average short-term strength. Integrating Equation (\ref{eq:usalpha0}) directly gives the failure time $T_s$  in terms of $A$, $B$, and $\mu_s$ as
\begin{eqnarray}\label{eq:ussolution}
T_s &=& \frac{\mu_s \cdot B \exp\{A\}}{\exp\{B\} - 1}
 \end{eqnarray}
and Equation (\ref{eq:usalpha2}) implies
\begin{eqnarray*}
\label{eq:usalpharamp}
  \alpha(t)
  &=&\frac{\int_0^{(t/\mu_s)}\exp{\{\frac{ B u}{T_s/\mu_s}\} }du }{\int_0^{(T_s/\mu_s)} \exp{\{ \frac{Bu}{T_s/\mu_s}\}}du} \nonumber\\
	&=& \frac{\exp{\{Bt/T_s\}} - 1}{\exp{\{B\}} - 1}.
 \end{eqnarray*}

  \textbf{Remarks:}
  \begin{enumerate}
  \setcounter{enumi}{3}
   \item Observe that for the ramp load test
  \begin{eqnarray*}\label{eq:usalpha0ramp}
 	\dot{\alpha}(0)
 	&=&[ \mu_s \exp\{A\}]^{-1} \\
    \dot{\alpha}(T_s)
  &=&\frac{B}{T_s}\frac{\exp{\{B\}}}{\exp{\{B\}} - 1}.
 \end{eqnarray*}
These equations provide some intuition on the role of the random effects $A$ and $B$.  The first equation shows that the initial rate of damage accumulation for a specimen relative to the population is governed by its random effect $A$;  for a small $A$ that rate is faster.  The second equation shows that $B$ controls the accumulation rate as the specimen approaches its failure time; for large $B$ that rate is faster, since $\frac{B \exp{\{B\}}}{\exp{\{B\}} - 1}$ is an increasing function of $B$.
  \end{enumerate}

\subsection{The Canadian Model}\label{sec:canadianmodel}
  This subsection provides another instance of the model
  in Equation (\ref{eq:piversion}) in what
  is now referred to as the ``Canadian model''
  \citep{foschiyao1986dol}. As
  originally specified it is given by
  \begin{equation}\label{eq:canadianmodel}
  \dot{\alpha}(t) = a [\tau(t) - \sigma_0 \tau_s]_+^b
  + c [\tau(t) - \sigma_0 \tau_s]_+^n \alpha(t)
  \end{equation}
  where $a$, $b$, $c$, $n$, $\sigma_0$ are log--normally distributed
  random effects, $\tau_s$ (psi) is the short term breaking strength,
  $\tau(t)$ (psi) is the applied stress at time $t$
  and $\sigma_0$ is the stress ratio threshold (the subscript $+$ indicating
  that the quantity in square brackets becomes $0$
  when the quantity inside those brackets is negative). As before, the conditions $\alpha(0)=0$ and $\alpha(T_l) = 1$ will determine $T_l$ for the given $\tau(t)$ as
a function of the specimen specific random effects.

As with the US model, the first step  nondimensionalizes time, by replacing the left hand side of Equation (\ref{eq:canadianmodel}) by $\pi_1(t)$.  Then the right hand side must be unitless as well; however, as formulated the units associated with both terms on the right hand side of the model involve powers, $b$ and $n$. These lead respectively to units in those terms of $(psi)^b$ and $(psi)^n$. But the coefficients, $a$ and $c$, cannot involve those random powers and so cannot compensate to make those two terms unitless.  A simple adjustment re--expresses that equation using $\pi_3(t) = \tau(t)/\tau_s$, so that
  \begin{eqnarray*}
 	\pi_1(t) &=& [(\tilde{a} \tau_s )(\tau(t)/\tau_s - \sigma_0)_+]^b
  + [(\tilde{c} \tau_s )(\tau(t)/\tau_s - \sigma_0)_+]^n \alpha(t)\\
  	&=& [(\tilde{a} \tau_s )(\pi_3(t) - \sigma_0)_+]^b
  + [(\tilde{c} \tau_s )(\pi_3(t) - \sigma_0)_+]^n \alpha(t),
  \end{eqnarray*}
  where $\tilde{a}$ and $\tilde{c}$ are now random effects with $[\tilde{a}] = [\tilde{c}] = F^{-1}L^2$.

  To illustrate the use of this model, again consider the special case of a ramp load test where $T_l = T_s$, $\tau_s = kT_s$, and $\pi_3(t) = t/T_s$. Then
  \begin{eqnarray}\label{eq:canramp}
 	\dot{\alpha}(t) \mu_s &=& [\tilde{a} k T_s(t/T_s - \sigma_0)_+]^b
  + [\tilde{c} k T_s (t/T_s - \sigma_0)_+]^n \alpha(t).
  \end{eqnarray}
As before, the loading rate $k$ is known and may be specimen specific.  Define the integrating factor
  \begin{eqnarray*}
  H(t) &=& \exp \left\{ \int -\frac{1}{\mu_s} \left[ \tilde{c} k T_s \left( \frac{t}{T_s} - \sigma_0 \right) \right] ^n \,dt \right\} \\
  &=& \exp \left\{ -\frac{1}{\mu_s}  (\tilde{c} k T_s)^n \frac{T_s}{n+1}  \left( \frac{t}{T_s} - \sigma_0 \right)^{n+1} \right\}.
  \end{eqnarray*}
  Then
  \begin{eqnarray*}
\frac{d}{dt} \left[ \alpha(t) H(t) \right]  = \frac{1}{\mu_s} \cdot H(t) \left[ \tilde{a} k T_s  \left( \frac{t}{T_s} - \sigma_0 \right) \right]^b.
  \end{eqnarray*}

  For this model no damage is accumulated until the stress ratio threshold reaches $t = \sigma_0 T_s$. Integration then yields
  \begin{eqnarray*}
  \alpha(T_s)H(T_s) - \alpha( \sigma_0 T_s)H(\sigma_0 T_s) = \int_{\sigma_0 T_s}^{T_s} \frac{1}{\mu_s} \cdot H(t) \left[ \tilde{a} k T_s  \left( \frac{t}{T_s} - \sigma_0 \right) \right]^b \,dt.
  \end{eqnarray*}
The change of variables $u = -\log H(t)$ and the positivity of $-\log H(t)$ yields
  \begin{eqnarray*}
H(T_s) = \frac{( \tilde{a} k T_s )^b }{ (\tilde{c} k T_s)^{n(b+1)/(n+1)} } \left(\frac{\mu_s (n+1)}{T_s}\right)^{\frac{b-n}{n+1}} \int_0^{-\log H(T_s)} e^{-u}  u^{(b+1)/(n+1) - 1}\, du,
  \end{eqnarray*}
the integral being the lower incomplete Gamma function, which can be evaluated numerically using standard mathematical libraries.  Given the values of random effects $a$, $b$, $c$, $n$, and $\sigma_0$, $T_s$ is determined by the solution to this equation.  Unlike the US model however, this equation does not have an analytical solution, and must be solved numerically.
\vskip .1in
  \textbf{Remarks:}
  \begin{enumerate}
  \setcounter{enumi}{4}
  \item Some later authors (e.g., \citet{kohler2002probabilistic} and  \citet{hoffmeyer2007duration}) state the Canadian model in the following manner instead:
  \begin{equation}\label{eq:canadianmodel2}
  \dot{\alpha}(t) = a \left( \frac{\tau(t)}{\tau_s} - \sigma_0 \right)_+^b
  + c \left( \frac{\tau(t)}{\tau_s} - \sigma_0 \right)_+^n \alpha(t).
  \end{equation}
  Note that this is an alternative way to resolve the inconsistent $psi$ units on the right hand side of Equation (\ref{eq:canadianmodel}).  However it also fundamentally changes the nature of the dependence of $\dot{\alpha}(t)$ on $\tau_s$.  In particular, fitting the model to ramp load data, using the specification in Equation (\ref{eq:canadianmodel2}) will not explicitly depend on the loading rate $k$, while the original Canadian model does.

  \end{enumerate}

  \section{The experiment}\label{sec:experiment}
The experimental data consists of $n=98$ specimens of 12-ft 1650f-1.5E Spruce-Pine-Fir (SPF) 2x4 randomly drawn from a bundle and tested destructively under short--term bending loads.  The bending machine was set up for a span corresponding to a span--to--depth ratio of 21:1 (73.5 inches) for testing in accordance with ASTM D 4761, Section 6-10 \citep{astm4761test}.
  The edge to be stressed in tension was selected randomly, and the maximum strength reducing characteristic was randomly located in the 73.5-inch test span.  Specimens were trimmed to remove the excess overhang, allowing for 4 inches past each end (a total length 81.5 inches).

\begin{figure}[!ht]
\begin{center}
\includegraphics[width=6.5in]{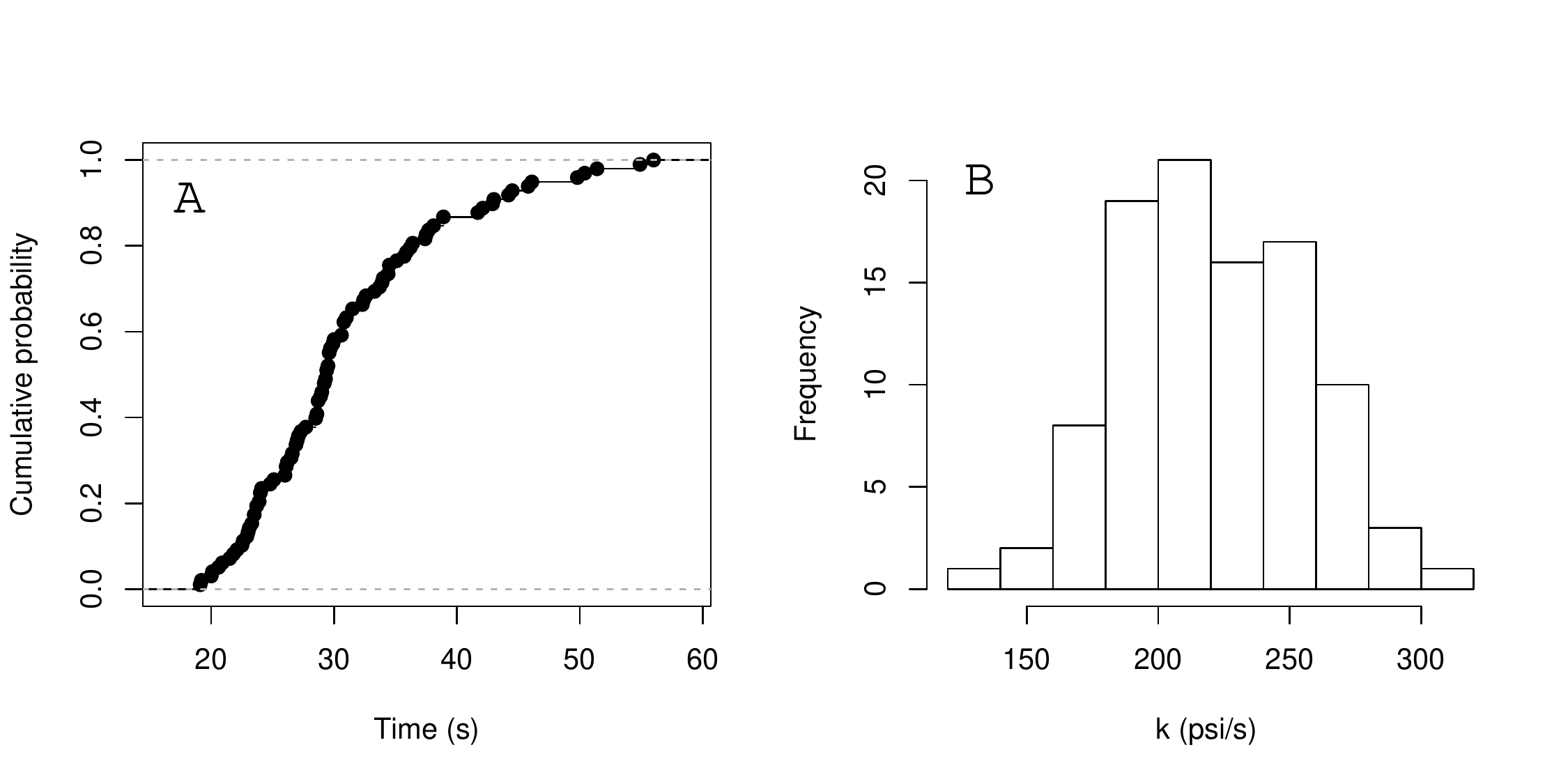}
\caption{(A) Empirical cumulative distribution of short-term bending failure times; (B) Ramp loading rate of specimens}
\label{fig:benddata}
\end{center}
\end{figure}

The load profile was set to a constant deflection rate of $d=0.045 in/s$.  This deflection rate translates to approximate ramp load tests with a specimen specific loading rate $k$ that depends on its elasticity $E$.  As shown in Equation (\ref{eq:deflectionrateload}), the loading rate $k$ is approximately linear in $E$, and thus the variability in $k$ can be attributed to the variability in $E$ among the specimens in the sample.  The time until failure ($T_s$) was recorded for each specimen.  Figure \ref{fig:benddata} shows the empirical cumulative distribution of $T_s$ and a histogram of the realized loading rates $k$ for the sample.  Based on these data, set the reference failure time $\mu_s$ to be the sample mean of $31.0$ seconds.

\section{Data analysis}\label{sec:results}

\subsection{Overview of Bayesian statistical methods}
Methods to fit the experimental data to the models discussed in Sections \ref{US:Model} and \ref{sec:canadianmodel} are now developed.  For the US model, assume that $A$ and $B$ in Equation (\ref{eq:ussolution}) are specimen specific random effects, log--normally distributed with parameters $(\mu_A, \sigma^2_A)$ and $(\mu_B, \sigma^2_B)$, respectively.  For the Canadian model, assume $\tilde{a}$, $b$, $\tilde{c}$, $n$, and $\sigma_0$ in Equation (\ref{eq:canramp}) are specimen specific random effects.  Assume as  in the original Canadian model's derivation, that the unitless random effects $b$ and $n$ are log--normally distributed, with respective parameters $(\mu_b, \sigma^2_b)$,  $(\mu_n, \sigma^2_n)$.  Since the stress ratio satisfies $0 < \sigma_0 < 1$, the Normal distribution may be adopted for $\sigma_0$ after a logit--transformation, with parameters $(\mu_{s0}, \sigma^2_{s0})$.  The remaining random effects $\tilde{a}$ and $\tilde{c}$ are problematical since they have units $F^{-1}L^2$, thus ruling out use of the log--normal distribution as was done in the past.  Instead the Normal distribution has been chosen as an approximation, with parameters $(\mu_a, \sigma^2_a)$ and $(\mu_c, \sigma^2_c)$, respectively, so that now these parameters can have the appropriate units.  In all cases, $\mu$ is the mean parameter and $\sigma^2$ refers to the variance parameter of the distribution.  Finally, while the theoretical failure times $T_s$ are deterministic solutions to equations involving the random effects, this condition may be relaxed to accommodate model error.

A Bayesian statistical approach is adopted for fitting these models \citep{gelman2014bayesian}.  As far as the authors know, such methods have not been previously applied to estimate parameters for ADMs, so a brief review is provided along with a description of their merit in the problem at hand.  Bayesian analysis combines two ingredients:  the `prior', which is a probability density specified on the parameters to represent the investigator's knowledge before the experiment is done; and the likelihood function of the observed data given parameter values.  The latter is the basis of the classical `maximum likelihood' approach to parameter estimation.  In the Bayesian setting, the product of the prior and likelihood gives the `posterior' distribution, which represents the probability distribution of the parameters after seeing the data.  That posterior is the basis of drawing conclusions about the parameters.

The advantages of a Bayesian approach for estimating ADMs are two-fold.  First, uncertainty about the parameters is  captured.  This is of particular importance since the random effects are not observed in the data; only the failure time $T_s$ and loading rate $k$.  Therefore, it is difficult to obtain reliable confidence intervals on the parameters by matching theoretical and empirical quantiles as was done in the past, see for example \citet{foschiyao1986dol}.  This is handled naturally in the Bayesian setting, since the posterior can be explored effectively using Markov Chain Monte Carlo (MCMC) techniques to provide genuine posterior probability intervals.  Second, the posterior can be used to construct predictive distributions for model checking or prediction for future specimens.  When the posterior is explored via MCMC simulation, such predictive distributions can be easily obtained by using the MCMC samples to numerically integrate out unknowns from the fitted model, as will be demonstrated.

\subsection{Analysis procedure}

In this subsection, the procedure to carry out the Bayesian analysis on the dataset and compare models is described.

The likelihood functions for both the US and Canadian models are first needed.  Let $\theta$ denote the vector of model parameters, and $T_i$, $\gamma_i$ denote the failure time and vector of unobserved random effects for specimen $i$ respectively.  In general, let $h(\gamma)$ denote the deterministically solved failure time corresponding to a random effect vector $\gamma$; since this solution does not readily yield a tractable likelihood function, an approximation is adopted by assuming that these solutions have accuracy to the nearest second for data on the current time scale ($\sim 30$s).  This assumption accommodates model error and implies that the observed $T_s$ lies uniformly randomly in the interval $[h(\gamma) - 0.5s, h(\gamma) + 0.5s]$.  Recall that the US solution is given analytically in Equation (\ref{eq:ussolution}), while the Canadian solution must be found numerically.

The general notation $p(a | b)$ is used to denote the probability distribution of $a$ conditional on $b$.  Then the likelihood $L$ of the parameters $\theta$ for specimen $i$ is
\begin{align*}
L(\theta| T_i) &= p(T_i | \theta) \\
&= \int p(T_i | \gamma_i) p(\gamma_i | \theta) \,d\gamma_i \\
&= \int  I\left[h(\gamma_i) - 0.5 \le T_i \le  h(\gamma_i) + 0.5 \right] \times  p(\gamma_i |\theta) \,d\gamma_i,
\end{align*}
where $I$ denotes the indicator function where $I(x) = 1$ if $x$ is true, and 0 otherwise.

This integral cannot be done analytically, but can be evaluated using Monte Carlo integration: draw $N$ realizations of $\gamma_i$ from its distribution given the current values of $\theta$, where $N$ is a large integer.   Denote these values by  $\gamma_i^{(1)},\gamma_i^{(2)}, \ldots, \gamma_i^{(N)}$.  Then a large enough $N$ yields a result arbitrarily close to the true likelihood value via the estimate
\begin{equation}\label{eq:likapprox}
L(\theta| T_i) \approx \frac{1}{N} \sum_{j=1}^N  I\left[h(\gamma_i^{(j)}) - 0.5 \le T_i \le  h(\gamma_i^{(j)}) + 0.5 \right],
\end{equation}
which in other words is simply the proportion of samples where the observed $T_i$ is compatible with $\gamma$.  Note that this likelihood does not have an analytical gradient, and the necessity of Monte Carlo integration in its calculation would render numerical gradients to be unstable.  Hence a direct maximization of the likelihood function (for a maximum likelihood analysis) is not straightforward, but this poses no difficulty for the MCMC techniques adopted here.

Next, priors must be specified on the parameters.  Here it is assumed, \emph{a priori}, that all the parameters in $\theta$ are statistically independent.  Let the $\mu$ parameters have a $Normal(0, 100^2)$ prior density and the $\sigma^2$ parameters have a $Inv$-$Gamma(0.001,0.001)$ prior density.  These choices of priors represent the absence of any prior knowledge on the parameters \citep{gelman2006prior}.  Then, assuming the test sample consists of $n$ statistically independent specimens, the posterior distribution of the parameters is given by
$$
p( \theta | T_1, \ldots, T_{n} ) \propto p(\theta) \prod_{i=1}^n p(T_i | \theta),
$$
where $p(\theta)$ denotes the joint probability density of the priors.

When the posterior is analytically intractable, as is the case here, inference can be made by drawing representative samples from this probability distribution using MCMC simulation techniques \citep{brooks2011handbook}.  The particular variant of MCMC used here for efficiency is parallel tempering \citep{swendsen1986replica} on the power posterior with Metropolis--Hastings iterations on each computing node.  Empirical assessments suggest that using $N=10,000$ draws in Equation (\ref{eq:likapprox}) provides sufficiently reliable calculations (absolute error in the log-posterior $< 1.0$).

In the Bayesian setting, model comparison is often carried out by calculating the Bayes Factor, to determine which model is more strongly supported by the data \citep{kass1995bayes}.  The Bayes Factor in favour of the Canadian Model ($M=1$) versus the US model ($M=2$) is defined as:
\begin{eqnarray}\label{eqn:bayesfactor}
B_{12} = \frac{p(T_1, \ldots, T_n | M = 1)}{p(T_1, \ldots, T_n | M = 2)},
\end{eqnarray}
where the term $p(T_1, \ldots, T_n | M = m)$ is known as the \emph{marginal likelihood} of model $m$.  The calculation of the marginal likelihood integrates out the model parameters, thus taking into account the model complexity and number of parameters.  Hence the Bayes factor, which is the ratio of the marginal likelihoods, directly evaluates which of the two models is more strongly supported by the data, with $B_{12} > 1$ indicating that model $M=1$ is more strongly supported by the data than $M=2$; $B_{12} > 150$ is generally considered as `very strong' or decisive evidence \citep{kass1995bayes}.  Here to calculate the numerical value of Equation (\ref{eqn:bayesfactor}) from the MCMC samples, Equation (7) in \citet{friel2008marginal} was used.

Finally, suppose an application of interest is to use model $m$ to predict the failure time $T_f$ for a specimen.  The Bayesian framework provides the probability distribution of $T_f$ as
\begin{eqnarray}\label{eqn:bayesprediction}
p(T_f | M = m) = \int p(T_f | \theta, M = m) p(\theta | T_1, \ldots, T_n, M=m) \,d\theta
\end{eqnarray}
This distribution can be applied to predict failure times of future specimens and to check the quality of the model fit on the existing data.  These are illustrated in the following section.

\subsection{Results}

The fitted US and Canadian models based on the experimental data are presented first.  Each computing node in the parallel tempering MCMC setup ran 10,000 Metropolis-Hastings iterations, with the first 1000 samples discarded as burn--in.  Table \ref{tab:fitted} summarizes the key quantiles from the resulting posterior distributions of the parameters.  Consider the 50\% quantile (median) to be the point estimate of each parameter; the 2.5\%  and 97.5\% quantiles can be interpreted as the endpoints of the 95\% Bayesian credible interval, i.e.~the posterior probability that the parameter lies within it is 0.95.  The marginal likelihoods of the two models (on the log-scale) are also shown; these yield the Bayes Factor $B_{12} = 6.8 \times 10^5$, and this magnitude of $B_{12}$ suggests the data strongly favours the Canadian model \citep{kass1995bayes} for this particular dataset.

\begin{table}[htbp]
    \centering
    \caption{Summaries of posterior distributions of parameters for the US and Canadian models.}
    \begin{tabular}{cc}
    \begin{tabular}[t]{rrrr}
     \multicolumn{4}{c}{\textbf{US Model}} \\
    & \multicolumn{3}{c}{Posterior quantiles} \\
        & 50\% & 2.5\% & 97.5\% \\
         \hline
     $\mu_A$ &   0.643 & -0.479 & 1.439 \\
     $\sigma_A$ &  0.100 & 0.020 & 0.356 \\
     $\mu_B$ &   1.15 & 0.25 & 1.81 \\
     $\sigma_B$ &  0.036 & 0.015 & 0.119 \\
     \hline
    \multicolumn{4}{l}{Marginal log-likelihood: -339.7}
    \end{tabular} &
        \begin{tabular}[t]{rrrr}
     \multicolumn{4}{c}{\textbf{Canadian Model}} \\
    & \multicolumn{3}{c}{Posterior quantiles} \\
        & 50\% & 2.5\% & 97.5\% \\
         \hline
     $\mu_a$ &   1.97 & 0.228 & 4.36 \\
     $\sigma_a$ &   0.0357 & 0.0162 & 0.33 \\
     $\mu_b$ &   1.84 & -1.56 & 4.13 \\
     $\sigma_b$ &   0.0741 & 0.0153 & 0.72 \\
     $\mu_c$ &   2.29 & 0.252 & 6.29 \\
    $\sigma_c$ &    0.0317 & 0.0147 & 0.551 \\
    $\mu_n$ &    -1.33 & -5.96 & 1.98 \\
     $\sigma_n$ &   0.0521 & 0.0162 & 0.931 \\
     $\mu_{s0}$ &   1.58 & -2.69 & 2.79 \\
     $\sigma_{s0}$ &   0.0435 & 0.0158 & 0.307 \\
     \hline
    \multicolumn{4}{l}{Marginal log-likelihood: -326.27}
    \end{tabular}
    \end{tabular}\label{tab:fitted}
\end{table}

\begin{figure}[!ht]
\begin{center}
\includegraphics[width=6.5in]{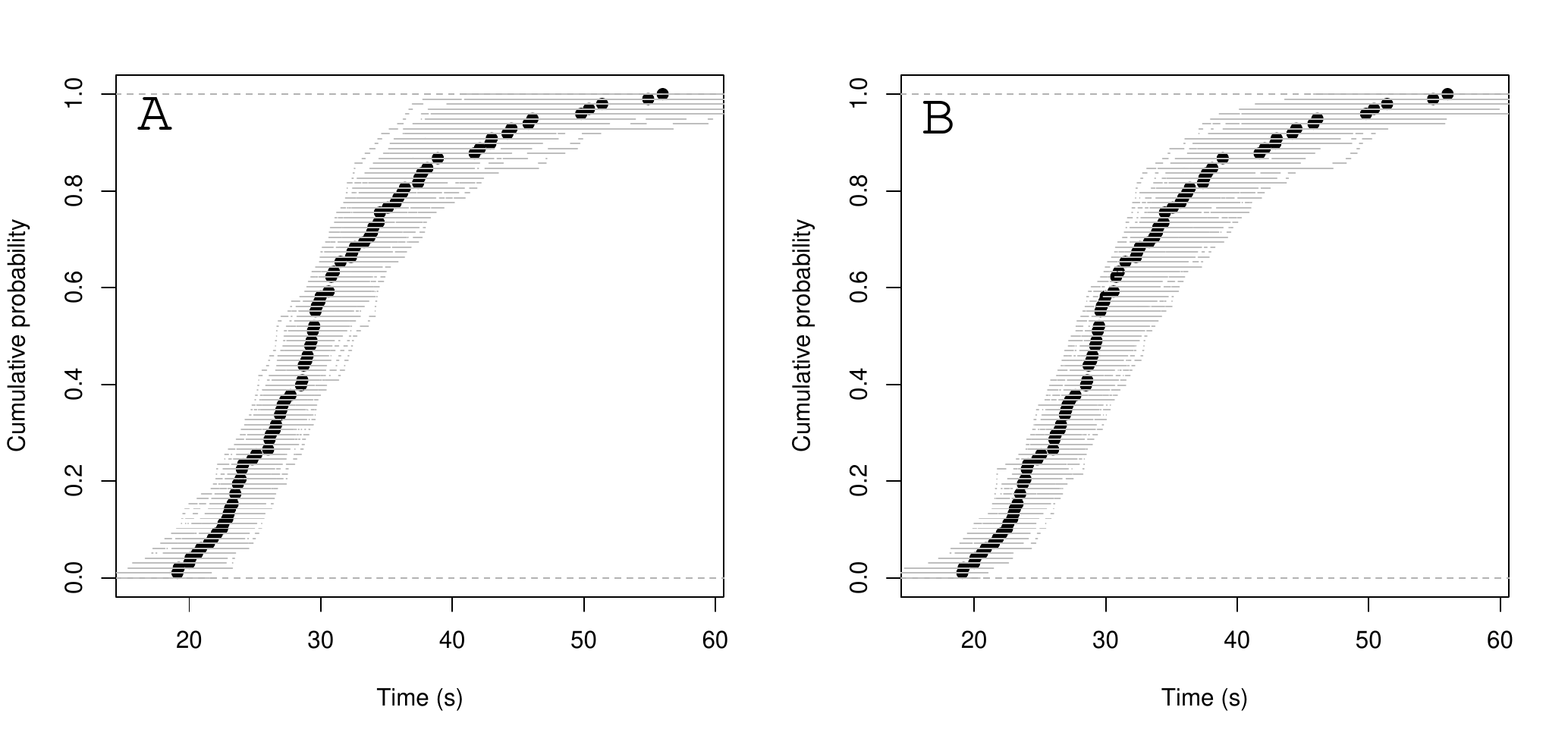}
\caption{Assessment of model fits via empirical cumulative distributions of data generated from fitted models.  (A) US Model, (B) Canadian Model.  It can be seen that the intervals for the Canadian model are narrower in both the lower and upper extremes of the distribution.}
\label{fig:bendrep}
\end{center}
\end{figure}

To assess how each model fits, Equation (\ref{eqn:bayesprediction}) was used to generate 100 hypothetical replicates of the dataset.  A visual of the fit quality is obtained by superimposing the empirical cumulative distributions of the replicates (in grey), onto the actual cumulative distribution of the data shown in Figure \ref{fig:benddata}.  The results are shown in Figure \ref{fig:bendrep}. Notice that while the central portions of the distributions appear to fit equally well, the Canadian model is better able to replicate the observed data in both the lower and upper extremes of the distribution.  There is less variability in the replicates (grey) around the observed data for the specimens with the shorter and longer failure times.

\begin{figure}[!ht]
\begin{center}
\includegraphics[width=6.5in]{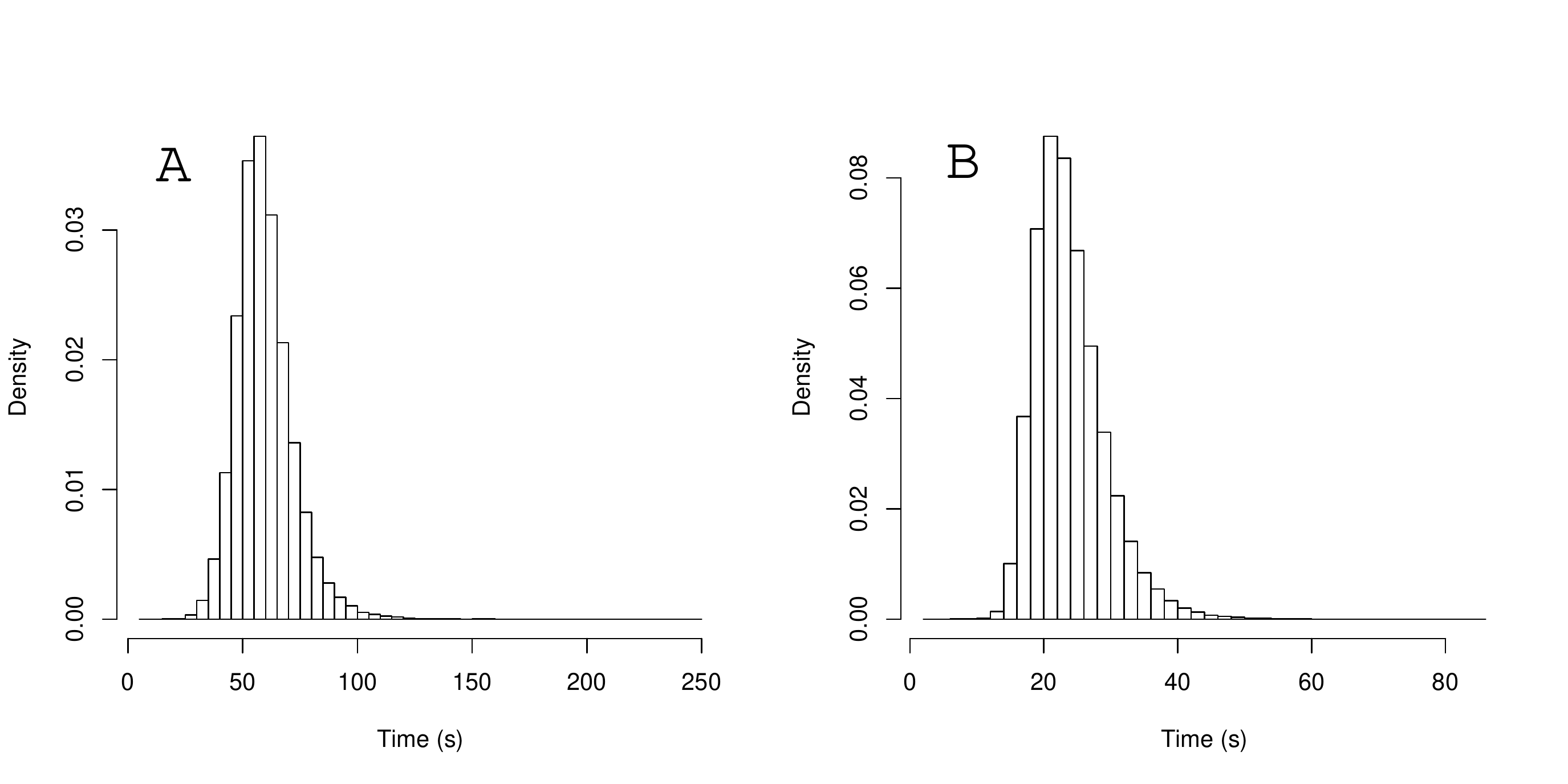}
\caption{Predictive distributions for the Canadian model using two different loading rates. (A) Loading rate $k = 0.1$, (B) Loading rate $k = 0.3$.  The average load sustained at failure is higher for the faster loading rate.}
\label{fig:pred}
\end{center}
\end{figure}

Finally Figure \ref{fig:pred} depicts two plots of predictive distributions based on the fitted Canadian model for this sample, again computed using Equation (\ref{eqn:bayesprediction}). For this purpose, two different ramp loading rates are compared:  the slower rate $k = 0.1$, and the faster rate $k = 0.3$.  These predictive distributions corroborate the expected effect: the group subject to the faster loading rate sustains a higher average load at failure.  The mean time to failure of the $k=0.3$ scenario is 23.8s, compared to 59.6s for the $k=0.1$ scenario.  These correspond to average loads at failure of 7158\emph{psi} and 5956\emph{psi}, respectively.

\section{Summary and concluding remarks}\label{sec:conclusions}
In this paper, a framework based on dimensional analysis was presented that enables one to build accumulated damage models.  The analysis in Section \ref{sec:model} shows the need to ensure dimensional coherence in model development. There two investigators, working on different time scales but using the very same accumulated damage model can reach strikingly different conclusions about the rate of damage accumulation. The problem for that model can be solved merely by changing it to recognize that as a transcendental function, $\exp{(x)}$, neither the function nor its argument $x$ can have units of measurement. But a deeper analysis, based primarily on an application of the celebrated Buckingham $\pi$ theorem \citep{buckingham1914physically}, ensures that the model does not depend on what scales are used for measurement.  The final result is a family of possible accumulated damage models from which to select a model in a specific application.

	The paper then explores how two well--known models -- the US Model and the Canadian model -- can be adapted to fit into that family while retaining their important features. The \emph{ad hoc} approach in Section \ref{sec:model} yielded conclusions about the  US model. But for the Canadian model, substantially more adaptation was needed.

The second major feature of this paper was a demonstration of how the resulting models could then be implemented within a Bayesian statistical framework in order to reflect all their associated uncertainties. That demonstration was carried out in the simplest case of ramp load testing using experimental data that the first author produced in an FPInnovations Vancouver laboratory. The empirical results for that dataset favoured the Canadian model.  In this case only one time scale for loading was considered, i.e.~an average failure time of $\sim$30 seconds, to show the merits of the Bayesian approach for working with these models.  The same statistical approach applies for analyzing data from different time scales.  Classic studies on rate--of--loading (e.g. \citet{karacabeyli1993rate}) have used ramp--load tests with different loading rates (e.g.~with average failure times set to 5 hours, 10 minutes, 1 minute, 1 second, etc.), as well as constant--load tests, to quantify the effect of load rate on strength.  That more extensive analysis will be the subject of follow-up work to this paper.

Overall the paper has provided a foundation for accumulated damage modelling on which can be used to build new models for setting design values for new engineered lumber products such as cross laminated timber or strand--based wood composites (see for example \citet{wang2012doltheory} and \citet{wang2012dolexperiment}).

One might well ask if such a foundation is needed. After all, the original Canadian model
did fit the experimental data rather well despite its dimensional inconsistencies -- that is, when implicitly the units of measurement were dropped.  The good fit is perhaps not surprising given the large number of  parameters in the model. One is reminded of John von Neumann's famous quip: ``With four parameters I can fit an elephant and with five I can make him wiggle his trunk.''

The authors' response would be that such models, which can only be fitted on accelerated test data, cannot be directly validated for their intended use in predicting long term reliability. Therefore they must be developed in accordance with good modelling practice, to ensure that they appear trustworthy. In particular the period of time until a piece of lumber fails does not depend on the units in which that period is measured, as ensured by application of the Buckingham $\pi$ theorem in this paper.  Another important feature of good modelling practice embraced in this paper is a method for fitting the model that comes with a characterization of the uncertainties associated with it.

Finally unpublished work by the authors done since the current paper was first submitted, based on differences in the way the analysis could be done as well as in the models, shows important differences in the results given by the analyses of the original Canadian model and the non--dimensionalized version presented here \citep{yang2017adm}.

\vskip .1in
{\bf Acknowledgements.} The authors are greatly indebted to Conroy Lum from FPInnovations for helpful discussions.  They are also indebted to FPInnovations and its technical support staff, for facilitating the experimental work that was done to produce the data used in this paper.  The work reported in this paper was partially supported by a Collaborative Research and Development grant from the Natural Sciences and Engineering Research Council of Canada.

\bibliographystyle{chicago}

\end{document}